%% file: main.tex
\documentclass[sigconf]{acmart}

\usepackage{color}

\usepackage{graphicx}
\usepackage{subfigure} 
\usepackage{float}

\usepackage{booktabs}
\usepackage{multirow}
\usepackage{color, colortbl}
\definecolor{LightCyan}{rgb}{0.88,0.88,0.88}
\definecolor{darkgreen}{rgb}{0.0, 0.5, 0.0}

\usepackage[toc,page]{appendix} 
\usepackage{tabularx}
\usepackage{url}
\usepackage{enumitem}
\setlist[itemize]{leftmargin=*}

\newcommand{\ly}[1]{{\color{black} #1}}

\usepackage[ruled]{algorithm2e}

\AtBeginDocument{%
  \providecommand\BibTeX{{%
    \normalfont B\kern-0.5em{\scshape i\kern-0.25em b}\kern-0.8em\TeX}}}

\copyrightyear{2025}
\acmYear{2025}
\setcopyright{cc}
\setcctype{by-nc-nd}
\acmConference[CHI '25]{CHI Conference on Human Factors in Computing Systems}{April 26-May 1, 2025}{Yokohama, Japan}
\acmBooktitle{CHI Conference on Human Factors in Computing Systems (CHI '25), April 26-May 1, 2025, Yokohama, Japan}\acmDOI{10.1145/3706598.3713933}
\acmISBN{979-8-4007-1394-1/2025/04}

\begin{document}

\title{``AI Afterlife'' as Digital Legacy: Perceptions, Expectations, and Concerns}

\author{Ying Lei}
\affiliation{
  \institution{Simon Fraser University}
  \city{Vancouver}
  \country{Canada}
}
\email{ying_lei@sfu.ca}

\author{Shuai Ma}
\authornote{Corresponding author.}
\orcid{0000-0002-7658-292X}
\affiliation{
  \institution{Aalto University}
  \city{Espoo}
  \country{Finland}
}
\email{shuai.ma@aalto.fi}

\author{Yuling Sun}
\affiliation{
  \institution{Fudan University}
  \city{Shanghai}
  \country{China}
}
\email{yulingsun.lv@gmail.com}

\author{Xiaojuan Ma}
\affiliation{
  \institution{The Hong Kong University of Science and Technology}
  \city{Hong Kong}
  \country{China}
}
\email{mxj@cse.ust.hk}

\renewcommand{\shortauthors}{Ying Lei, et al.}

\begin{abstract}
The rise of generative AI technology has sparked interest in using digital information to create AI-generated agents as digital legacy. These agents, often referred to as ``AI Afterlives'', present unique challenges compared to traditional digital legacy. Yet, there is limited human-centered research on ``AI Afterlife'' as digital legacy, especially from the perspectives of the individuals being represented by these agents.
This paper presents a qualitative study examining users' perceptions, expectations, and concerns regarding AI-generated agents as digital legacy. 
We identify factors shaping people's attitudes, their perceived differences compared with the traditional digital legacy, and concerns they might have in real practices. We also examine the design aspects throughout the life cycle and interaction process. 
Based on these findings, we situate ``AI Afterlife'' in digital legacy, and delve into design implications for maintaining identity consistency and balancing intrusiveness and support in ``AI Afterlife'' as digital legacy.
\end{abstract} 




\maketitle

\input{sections/01-Introduction}
\input{sections/02-RelatedWork}
\input{sections/03-Methodology}
\input{sections/04-Result}
\input{sections/05-Discussion}
\input{sections/06-Limitations_and_Future_Work}
\input{sections/07-Conclusion}

\begin{acks}
The work described in this paper was partially supported by a grant from the Research Grants Council of the Hong Kong Special Administrative Region, China (Project Reference Number: AoE/E-601/24-N).
\end{acks}


\bibliographystyle{ACM-Reference-Format}
\bibliography{main}

\appendix
\section{Thematic map with intermediary codes}
\label{sec: thematic map}
\begin{figure*}[htbp]
	\centering 
	\includegraphics[width=0.83\linewidth]{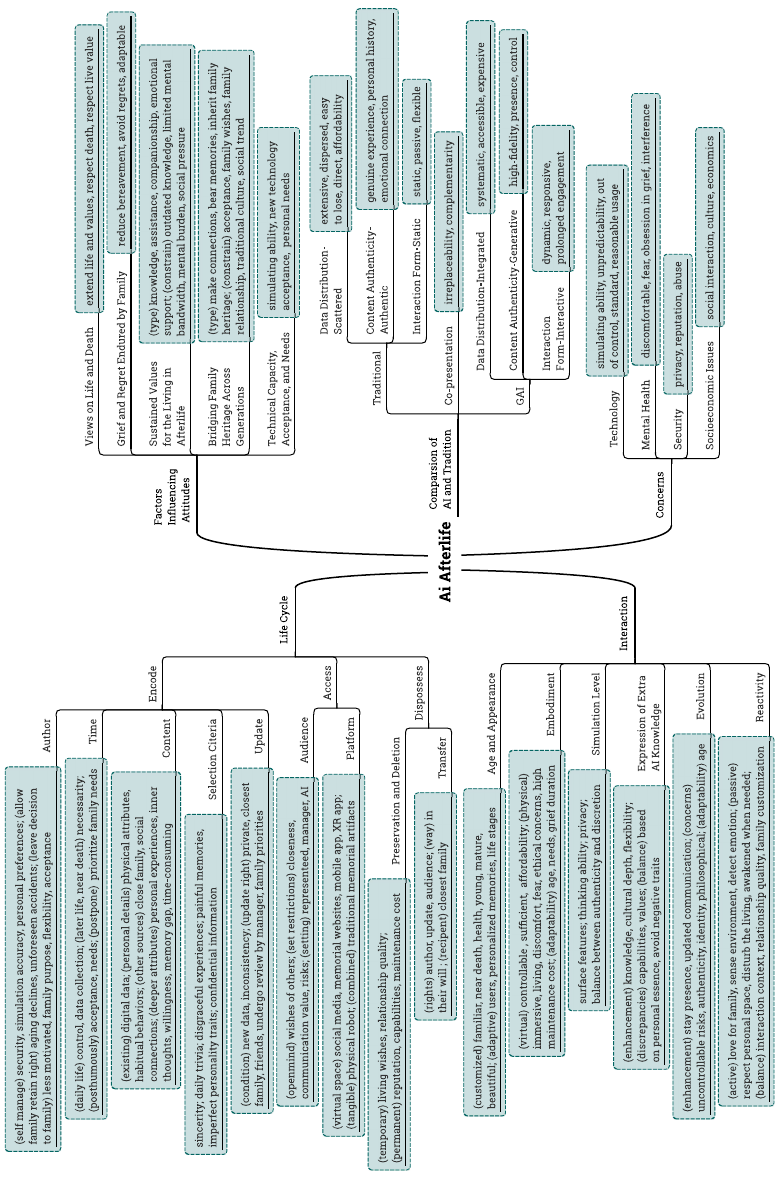}
    \caption{Thematic map with  intermediary codes.}
	\label{fig:thematic map}
        \Description{}
\end{figure*}

\end{document}

%% file: sections/01-Introduction.tex
\section{Introduction}

As generative AI technology rapidly advances in simulating specific individuals \cite{achiam2023gpt, anil2023palm, abramson2020creating, Eren_Coqui_TTS_2021, zhang2023sadtalker, la2024open, pataranutaporn2023living}, there is growing interest in using digital information to create AI-generated agents as digital legacy \cite{morris2024generative, brubaker2024ai}. 
We define this kind of agent, simulating individuals for post-mortem purposes, broadly as \textbf{\textit{``AI Afterlife''}}, which is expected to become commonplace within our lifetimes \cite{morris2024generative}. 
\label{definition of AI Afterlives}
``AI afterlives'', similar to other collections of digital legacies of the deceased, such as social media accounts \cite{ohman2019dead, gach2021getting}, passwords \cite{holt2021personal,ferreira2011password}, and files and media \cite{pfister2017will}, have the potential to capture meaningful aspects of a person’s life \cite{hunter2008beyond, gach2021getting}, ensure that loved ones can maintain access to significant parts of the deceased's life \cite{jung2023bereaved, kim2024maintaining, brubaker2024ai}, and help prevent potential distress or loss \cite{chen2021happens}.
\ly{This emerging form of digital legacy also has its unique benefit given the generative, interactive, and dynamic features of generative AI. They allow people to engage with the world even after death \cite{hunter2008beyond, gach2021getting} and to interact and maintain a healthy ``continuing bond'' with their loved ones \cite{jung2023bereaved, kim2024maintaining, brubaker2024ai}, especially during destabilizing and painful life changes \cite{gulotta2017digital, brubaker2024ai, morris2024generative, gulotta2014legacy, xygkou2023conversation}. }

Over the past few decades, Human-Computer Interaction (HCI) and Computer-Supported Cooperative Work (CSCW) communities have studied the concept of digital legacy, uncovering broader societal concerns about preserving and passing down digital legacy. Research in this area generally \ly{focuses} on the static digital legacy left by the deceased and can be categorized into four themes \cite{doyle2023digital}: digital legacy as reflections of identity (e.g., \cite{gulotta2014legacy, kaye2006have, thomas2014older}), people's engagement with digital legacy (e.g., \cite{gach2020experiences, massimi2011matters, brubaker2014stewarding, brubaker2016legacy}), practices for and implications of laying digital legacy to rest (e.g., \cite{gulotta2016engaging, kirk2010human, jamison2016ps}), and integrating technology into traditional legacy practices (e.g., \cite{odom2012technology, uriu2021floral, uriu2021generating}).

Despite that many of the existing findings on traditional digital legacy may be generalized to ``AI afterlife'', this new form of digital legacy has its unique challenges. \ly{First, the fact that it can continue to generate new information, react to people and the environment, and even evolve over time is likely to induce new answers to research questions under the aforementioned four themes \cite{doyle2023digital, brubaker2024ai, morris2024generative}.} In particular, the absence of common norms around digital legacy \cite{pfister2017will} poses a challenge that AI is likely to exacerbate \cite{morris2024generative}.
Second, while there are concrete practices \cite{guardian2024chinese, bbc2024the, cnn2024when, projectdecember, meet2020} for and research efforts \cite{abramson2020creating, morris2024generative, brubaker2024ai, AugmentedMIT} on creating AI-generated agents of deceased individuals by those who have lost a loved one \cite{guardian2024chinese, bbc2024the, cnn2024when}, more and more people are interested in joining the creation and governance of their own \ly{AI-generated agent-based digital legacies} while they are still alive \cite{hereafter2022, rememory}. 
There, however, has been little exploration from the perspectives of these individuals \cite{xygkou2023conversation}. 
Specifically, it is unclear how people perceive \ly{this form of digital legacy} when designing their own ``AI afterlives'', what their expectations are for the design and usage, and what concerns they might have in real practices. As digital legacy and generative AI become increasingly prevalent in both practice and research, seeking answers to these questions is crucial for researchers, designers, and developers to better address potential issues in integrating generative AI into existing digital legacy practices. Without such user-centered research, the development of AI-generated agent-based digital legacy may face legal, social, and ethical challenges \cite{guardian2024george, weidinger2023sociotechnical}.

This paper aims to address these gaps by providing an in-depth and empirical understanding of AI-generated agents as digital legacies, particularly from the perspectives of individuals being represented by these agents. We explore the nuanced perceptions, expectations, and concerns that individuals might have regarding their own ``AI Afterlives'' through semi-structured interviews with 18 participants from diverse demographic backgrounds.

Through a thorough thematic analysis, we identified several critical themes, including key factors influencing individuals' attitudes - encompassing personal, familial, technological, and societal aspects, as well as perceived differences between AI-generated agent-based and traditional digital legacy. Our findings also outline the life cycle of AI-generated agent-based digital legacy, detailing stages from encoding to accessing and dispossession. Additionally, \ly{we explore the desired interaction design for and the concerns participants have regarding these agents}. \ly{Our findings suggest the importance of maintaining identity consistency and balancing intrusiveness and support in ``AI Afterlife'' as digital legacy. }

Our study contributes to the HCI community by providing
(1) an empirical study of how individuals represented by AI-generated agents perceive and express concerns about ``AI Afterlife'' as digital legacy, 
(2) a user-centered speculative design to the life cycle and interaction of AI-generated agents in the context of digital legacy, and
(3) an in-depth discussion around ``AI Afterlife'' as digital legacy, alongside a set of design implications.

%% file: sections/02-RelatedWork.tex
\section{Related Work}


\subsection{Digital Legacy}

Legacy material refers to items passed down by the deceased to recipients such as bereaved loved ones \cite{doyle2023digital}. These materials include various aspects: the spirit aspect (e.g., values, wishes, identities), tangible items (e.g., objects, heirlooms), and even digital content \cite{doyle2023digital}. Such materials carry meanings that reflect important aspects of the deceased's life, allowing their legacy to extend beyond physical presence \cite{hunter2008beyond, gach2021getting}. They also ensure that bereaved loved ones maintain access to significant parts of the deceased's life \cite{jung2023bereaved, kim2024maintaining, brubaker2024ai}. This access is especially meaningful during emotional, spiritual, and logistical dilemmas surrounding the death of a loved one \cite{doyle2023digital}, thereby preventing potential distress or loss \cite{chen2021happens}. 

As many of the activities we do are now almost exclusively on digital devices, digital legacy, as valued collections of digital information, including social media accounts \cite{ohman2019dead, gach2021getting}, passwords \cite{holt2021personal, ferreira2011password}, files and media \cite{pfister2017will}, is becoming a more and more important part of legacy materials \cite{doyle2023digital, chen2021happens}.
However, preparing and passing down the digital legacy runs into numerous challenges (e.g., vast quantity and selection \cite{bergman2016science, vitale2018hoarding, jones2016curating}, cross devices and platforms \cite{odom2012lost, vertesi2016data}, limited time \cite{hunter2005leaving}, complex social environment \cite{doyle2023digital}, security and privacy \cite{edwards2013protecting, harbinja2019emails, holt2021personal}, lifespan \cite{gulotta2013digital}). These concerns have profound implications for researchers and designers to consider the potential of technologies in supporting digital legacy issues.

To assist in the design and deployment of effective technologies, over the past decades, HCI and CSCW researchers have conducted important research on preserving and passing down digital legacy. Doyle and Brubaker's recent literature review \cite{doyle2023digital} identified four key areas of focus within this field of research: how identity is navigated in the passing of digital legacy (e.g., \cite{gulotta2014legacy, kaye2006have, thomas2014older}), how digital legacy is engaged with (e.g., \cite{gach2020experiences, massimi2011matters, brubaker2014stewarding, brubaker2016legacy}), how digital legacy is put to rest (e.g., \cite{gulotta2016engaging, kirk2010human, jamison2016ps}), and how technology interfaces with offline legacy technologies (e.g., \cite{odom2012technology, uriu2021floral, uriu2021generating}). Much of this work has focused on static digital traces left behind after death \cite{morris2024generative}, such as social media profiles \cite{brubaker2016legacy, brubaker2014stewarding, brubaker2011we, brubaker2019orienting, gach2020experiences, gach2021getting, getty2011said, mori2012design}, personal archives \cite{kaye2006have, she2021living}, and burner accounts \cite{gulotta2016engaging}, which are central to legacy crafting and memorialization \cite{morris2024generative}. 

However, while traditional digital legacy offers valuable resources for reflection, this kind of digital legacy often lacks interactive, expressive, or evolving feature, failing to fully address people's desires to maintain agency over their posthumous future \cite{gulotta2017digital, brubaker2024ai, morris2024generative}. Meanwhile, their inability to evolve over time limits their capacity for deeper engagement and dynamic representation of an individual's identity, leaving gaps in continuity and interaction (e.g, continuing their life and passing on values \cite{gulotta2017digital}, and maintaining potential legal or economic benefits \cite{brubaker2024ai, morris2024generative}). Moreover, it has been shown that the simulation of deceased individuals can act as a buffer space and add resilience to loss in these destabilizing and painful life-changing moments for the bereaved \cite{xygkou2023conversation}, which can not be provided by traditional static digital legacy. 

To better accommodate people's needs in the afterlife, a more expressive, interactive, and evolving technology for creating digital legacy is needed. An emerging form of digital legacy, based on AI-generated agents, with its unique benefits in generative, interactive, and dynamic features suits these needs. Our study adds to this literature by adopting a user-centered method to explore the potential of AI-generated agent-based digital legacy.

\subsection{AI Afterlife}
Recent years have witnessed remarkable advancements in generative AI, especially large language models (LLMs) \cite{achiam2023gpt, touvron2023llama, touvron2023llama2, anil2023palm}, with progress made in generative images \cite{qu2023layoutllm}, video \cite{openai2024sora}, audio \cite{ghosal2023text}, and multi-modal models integrating various media types \cite{yu2024spae, girdhar2023imagebind}. These innovations have paved the way for AI-generated agents that can simulate realistic humans \cite{abramson2020creating}, including their voice \cite{Eren_Coqui_TTS_2021}, appearance \cite{musetalk, zhang2023sadtalker}, and even their personalities \cite{la2024open}, memories \cite{pataranutaporn2023living}, and so on. In particular, \ly{AI-generated agents and related technologies} are expected to gain fidelity and popularity as model capabilities improve and computational costs decrease \cite{hu2021lora,zheng2024response}, resulting in potential transformations in various fields (e.g., working \cite{brachman2024knowledge, ibmdigital}, education \cite{kazemitabaar2024codeaid,extance2023chatgpt}, healthcare \cite{yang2024talk2care}, entertainment \cite{isaza2024prompt}). 

As generative AI technology quickly advances in creating powerful and realistic agents of specific people, there is growing interest in creating AI-generated agents for deceased individuals by the bereaved loved ones, in order to continue bonds \cite{morris2024generative}. For instance, there have been concrete practices \cite{guardian2024chinese, bbc2024the, cnn2024when, projectdecember}, and research \cite{abramson2020creating, morris2024generative, brubaker2024ai} aimed at providing opportunities for the living ones to engage in conversations, seek comfort, or guidance from these agents of the deceased individuals. 
\ly{These AI-generated agents, simulating specific individuals for post-mortem purposes with various potential forms (e.g., chatbots \cite{bbc2024the, xygkou2023conversation}, avatars in mixed reality \cite{meet2020}, and robots \cite{berightback2013}), are called ``AI Afterlives'' in this paper (similar to ``generative ghosts'' in \cite{morris2024generative})}. They are anticipated to become common within our lifetimes for people to interact with loved ones and the broader world after death \cite{morris2024generative}. 

Meanwhile, more and more people are interested in joining the creation and governance of their own \ly{AI-generated agent-based digital legacies} while they are still alive \cite{hereafter2022, rememory}, with hopes of extending their influence and leaving a lasting impression \cite{gulotta2017digital, brubaker2024ai, morris2024generative}. Yet, there is little exploration from the perspectives of these people. In particular, it is unclear how people perceive it, what their expectations are for the design and usage, and what concerns they might have in real practices. Seeking answers to these questions is crucial for researchers, designers, and developers to better address potential issues in integrating generative AI into digital legacy, and avoid legal, social, and ethical challenges \cite{guardian2024george, weidinger2023sociotechnical}.

To better unpack the potential of using AI-generated agents to empower digital legacy, perspectives from users, especially the individuals represented by these agents, need more attention. Our study adds to this literature by investigating the perceptions, expectations, and concerns of AI-generated agent-based digital legacy.

%% file: sections/03-Methodology.tex
\section{Method} 
In this section, we outline the methods used in the design and execution of our qualitative study. 
Throughout the study, we apply the definition of ``AI Afterlife'' (mentioned in \autoref{definition of AI Afterlives}) consistently, while encouraging participants to imagine hypothetical post-mortem scenarios and technical forms of this emerging form of digital legacy.

\subsection{Participants: Inclusion Criteria and Recruitment}

\ly{Digital legacy is used for post-mortem purposes} \cite{doyle2023digital}. Therefore, we sought participants who met one or more of the following criteria: (1) type 1 - those who have had a near-death experience (e.g., due to illness, accidents, or aging), (2) type 2 - those who have experienced bereavement, and (3) type 3 - those who have deeply contemplated life and death.

We recruited participants primarily through personal social networks, social media groups, and snowball sampling \cite{goodman1961snowball}. The recruitment process began with the personal networks of two authors 
and outreach through relevant social media groups. Participants were then encouraged to recommend family members or friends who met our criteria, following the snowball sampling method. To ensure diversity, we aimed to recruit participants with varied demographic characteristics, including gender, age, health status, education, location, occupation, AI literacy, and related experiences.

\ly{After obtaining the IRB approval, we recruited participants for our qualitative study until we reached saturation, ensuring that no new themes or insights emerged from additional interviews \cite{guest2006many}.} We ultimately interviewed 18 participants, whose demographic details are presented in \autoref{tab:demographicinfo}. Among them, 11 were female and 7 were male, with ages evenly distributed between 20 and 69. They resided in 11 different provinces in China and represented a variety of occupational backgrounds. Their self-reported health statuses, assessed AI literacy levels, and experiences related to our topic were also diverse, further ensuring that we captured perspectives from individuals with different backgrounds.
\ly{To clarify, none of our participants had prior experience creating their own AI-generated agents. Instead, they contributed hypothetical perspectives on the potential applications and implications of this emerging form of digital legacy. }

\begin{table*}[htbp]
  \caption{Demographic information of participants. Rel. exp. means participants' related experience with death. Corp. Manager means the occupation - Corporate Manager. Type 1 means those who have had a near-death experience (e.g., due to illness, accidents, or aging); type 2 means those who have experienced bereavement; type 3 means those who have deeply contemplated life and death. }
  \label{tab:demographicinfo}
  \begin{tabular}{lllllllll}
    \toprule
    ID & Gender & Age & Location & Health & Education & Occupation & AI Literacy & Rel. exp. \\
    \midrule
    P1 & Male & 20-29 & Hongkong & Fair & Master's & Student & Advanced & type1, 3 \\
    P2 & Female & 40-49 & Hunan & Moderate & High School & Business Owner & Beginner & type1, 2 \\
    P3 & Female & 50-59 & Chongqin & Fair & Bachelor's & Teacher & Beginner & type2, 3 \\
    P4 & Female & 40-49 & Shandong & Healthy & Associate's & Bank Staff & Novice & type3 \\
    P5 & Female & 50-59 & Henan & Moderate & Master's & Lecturer & Intermediate & type1-3 \\
    P6 & Female & 20-29 & Beijing & Healthy & Master's & Student & Intermediate & type2, 3 \\
    P7 & Male & 20-29 & Shandong & Healthy & Bachelor's & Student & Advanced & type3 \\
    P8 & Male & 50-59 & Henan & Fair & Bachelor's & Corp. Manager & Advanced & type2, 3 \\
    P9 & Female & 40-49 & Beijing & Fair & Doctorate & Professor & Intermediate & type2, 3 \\
    P10 & Male & 30-39 & Liaoning & Healthy & Master's & Doctor & Novice & type2, 3 \\
    P11 & Female & 30-39 & Tianjin & Healthy & Master's & UX Researcher & Advanced & type2, 3 \\
    P12 & Male & 30-39 & Beijing & Healthy & Doctorate & Professor & Advanced & type2, 3 \\
    P13 & Female & 60-69 & Guangdong & Moderate & Bachelor's & Corp. Manager & Beginner & type1 \\
    P14 & Female & 50-59 & Fujian & Healthy & Associate's & Designer & Novice & type1-3 \\
    P15 & Female & 60-69 & Guangdong & Poor & High School & Plant Worker & Novice & type1-3 \\
    P16 & Male & 60-69 & Chongqing & Poor & Middle School & Plant Worker & Beginner & type1-3 \\
    P17 & Male & 70-79 & Guangdong & Moderate & Middle School & Trader & Intermediate & type1-3 \\
    P18 & Female & 70-79 & Beijing & Fair & Bachelor's & Corp. Manager & Beginner & type1-3 \\
  \bottomrule
\end{tabular}
\end{table*}

\subsection{Data Collection: Semi-structured Interviews}
We conducted semi-structured interviews to gain a deep understanding of participants' perceptions, expectations, and concerns regarding the use of ``AI Afterlife'' as digital legacy. 
\ly{At the beginning of each interview, we provided a brief introduction to legacy, digital legacy, AI, generative AI, and existing practices of using this new technology for post-mortem purposes. This introduction aimed to give participants a clear understanding of the social and technical background of our study. We then presented the definition of ``AI Afterlife'' in our study (mentioned in \autoref{definition of AI Afterlives}), emphasizing the flexibility in imagining digital forms to clarify our research focus and scope. In particular, we highlighted that the purpose of ``AI Afterlife''} is limited to creating their digital legacies in post-mortem contexts, that is, the expected simulated entities are the representations of participants themselves, not their deceased loved ones.

The interview questions were designed based on the digital legacy literature \cite{doyle2023digital, morris2024generative} and organized into three main parts. The first part focused on participants' daily practices with traditional digital information (e.g., digital photos and social media accounts), their attitudes towards using AI-generated agents as their digital legacies, and their perceptions of the relationships and differences between traditional digital legacy and AI-generated legacy. The second part explored participants' design expectations for AI-generated agents as digital legacies, covering aspects such as the life cycle of the agents and interaction design. The third part addressed participants' concerns about ``AI Afterlife'' as digital legacy, including potential ethical, legal, and social issues. 
The questions were designed to be inclusive, accommodating interviewees with varying levels of experience. When interesting points or relevant experiences were mentioned, we followed up with additional questions to gather more details and concrete examples.

The interviews were conducted remotely in Mandarin between June and August 2024 by two authors via Tencent Meeting, WeChat calls, or phone calls. \ly{Each interview lasted about 60 minutes with 15 USD compensation, allowing for an in-depth exploration of participant perspectives.} All interviews were recorded and transcribed verbatim in Chinese. 

\subsection{Data Analysis: Thematic Analysis}

We applied thematic analysis \cite{braun2012thematic} to analyze all the interview data using an inductive approach \cite{patton1990qualitative}. This method, grounded in the data itself, allowed us to explore emerging themes organically, ensuring flexibility and alignment with our research objectives. Two of the authors were involved in the data analysis process. First, we used a speech-to-text tool to transcribe the recorded interviews, which were then manually reviewed and corrected by the two authors.

We began coding and analysis concurrently with data collection, 
adopting an iterative process to identify and refine emerging themes. Manual coding was performed using Google Documents and Sheets. During the open coding phase, both authors independently reviewed the data and generated initial codes related to our research questions. We held weekly meetings to discuss and reconcile these codes, establishing a consensus to apply consistent codes for the same passages. This process enhanced coding consistency among researchers while maintaining alignment with the study objectives \cite{creswell2017research}.
\ly{
We continuously assessed thematic saturation throughout the interview process, conducting interviews in batches of three participants. In total, we conducted six batches. Thematic saturation was determined to be achieved when no new themes emerged, as observed in the most recent batch of interviews \cite{guest2006many}. }
After finalizing the initial code list, we refocused our analysis at a broader thematic level, synthesizing related codes into overarching themes. Through several rounds of discussion and refinement, we developed a thematic map that encapsulated five primary themes, \ly{as shown in \autoref{fig:thematic map} in \autoref{sec: thematic map}, which were detailed in \autoref{result} with representative quotes translated from Chinese to English.}

\subsection{Ethical Considerations}

We obtained ethical approval from the Ethics Committee of the authors' institution to conduct all procedures involving human subjects. Throughout the research process, we took careful steps to protect participants' rights and privacy. We also sought feedback from peers both within and outside our faculty to validate the ethical aspects of our research and to refine our interview procedures and questions, minimizing the risk of discomfort for participants. Before conducting the interviews, we fully informed participants of our intentions and provided background information, assuring them that the data would only be used for research. \ly{We obtained their informed consent before proceeding and clearly stated that they could withdraw at any time if they felt uncomfortable. All collected data were anonymized, with participants denoted as PX and no identifiable links to their identities.}

%% file: sections/04-Result.tex
\section{Result} \label{result}
In alignment with our research questions, we present the key findings in six dimensions, consisting of predefined themes based on the existing literature and emerging themes identified through thematic analysis: (1) factors influencing attitudes, (2) comparing AI-generated legacy and traditional digital legacy, (3) life cycle, (4) interaction, and (5) concerns. 

\subsection{Factors Influencing Attitudes: Person, Family, Technology and Society}\label{attitudes factors}

The attitudes of our participants towards the ``AI Afterlife'' as digital legacy are diverse and complex, being shaped by personal, familial, technological, and social factors, as shown in \autoref{tab:factors influencing attitudes}.
Generally speaking, personal views on life and death, along with concerns about the grief and loss of loved ones, significantly influence their perspectives. Furthermore, the desire to create lasting values and preserve family heritage can increase interest in ``AI Afterlife'', although this is also affected by family needs, technological capabilities, and broader social circumstances.

\begin{table*}[htbp]
  \caption{Summary of themes and main points in factors influencing attitudes on ``AI Afterlife'' as digital legacy.}
  \label{tab:factors influencing attitudes}
  \begin{tabular}{p{0.2\textwidth}p{0.75\textwidth}}
    \toprule
    \textbf{Theme} & \textbf{Main Point} \\
    \midrule
    \multirow{2}{=}{Views on Life and Death} 
            & Serve as a means of extending life and values (P1-3,11-14,17) \\ 
             & \cellcolor{gray!15} Blur the line between life and death (P4-5,7-10,15) \\
    \hline
     \multirow{3}{=}{Grief and Regret Endured by Family} 
            & Alleviate the grief of their loved ones (P2-3,6,12,14) \\ 
             & \cellcolor{gray!15} Prepare their loved ones for facing similar regrets (P2-4,11,15-16) \\ 
             & Attitudes varied with specific circumstances (P2,4,9) \\
    \hline
    \multirow{5}{=}{Sustained Values for the Living in Afterlife} 
            & \cellcolor{gray!15} Pass on practical knowledge and offer assistance (P1,3,13,14) \\ 
            & Provide daily companionship and emotional support (P6-7,10,13) \\ 
            & \cellcolor{gray!15} Lack relevance over time due to outdated knowledge (P8,15,17) \\ 
            & Limited emotional bandwidth to engage with them amidst busy lives (P5,8) \\
            & \cellcolor{gray!15} Impose mental burden and social pressure, disrupting daily life (P4, 8, 10, 12, 17-18) \\ 
    \hline
     \multirow{3}{=}{Bridging Family Heritage Across Generations} 
             & Maintain and share cherished family memories (P5,14,16,18) \\ 
             & \cellcolor{gray!15} Hesitant due to uncertainties about family dynamics (P7-8,15-16,18) \\ 
             & Influenced by evolving social trends (P15-18) \\ 
    \hline
    \multirow{3}{=}{Technical Capacity, Acceptance, and Needs} 
             & \cellcolor{gray!15} Simulation quality of human emotion and thought (P1,8,17) \\ 
             & Traditional attitudes toward new technologies (P11,14,16-18) \\ 
             & \cellcolor{gray!15} Match the features of a specific technology with personal needs (P1-2,7-8,14,17) \\
    \bottomrule
  \end{tabular}
\end{table*}

\subsubsection{Views on Life and Death.} \label{Views on Life and Death}

\ly{Participants' perceptions of life and death impact their acceptance of ``AI Afterlife'' as digital legacy. }
\ly{Those with positive opinions emphasized the spiritual aspect of life over the physical aspect of death, and viewed ``AI Afterlife'' as a means to extend their life and values (P1-3,11-14,17),} with P3 and P12 both expressing, \textit{``As long as I am remembered, my life continues.''} 
\ly{From their perspective, ``AI Afterlife'' provides opportunities to enhance the remembrance of both their image and spirit through more interactive and engaging connections with the living. }
P17 regarded experiences as the most valuable assets, and wished, \textit{``The emergence of digital human allows my experiences to be re-told and my life to leave traces in the lives of my children.''}

\ly{In contrast, participants with more reserved attitudes adhered to a strict definition of the boundary of death, focusing on its physical aspect and highlighting the natural cycle of life (P4-5, 7-10, 15).}
Just like the metaphor P15 made, \textit{``My life is like the burning of a candle, and my death is like the extinction of it.''}
\ly{However, ``AI Afterlife'', which allows people to engage with the world after death, might blur the line between life and death, challenging the traditional understanding of death as a natural endpoint.}
\ly{For example, P5 noted it might diminish the value of actual living,} \textit{``If my value continues to be highlighted through AI-generated agents after my death, what is the meaning of my biological life?''} 
\ly{Similarly, P10 expressed his concern on the potential consequence of hindering natural grieving and lessening the value of cherishing loved ones in life,} \textit{``The pain caused by the death of loved ones is part of the life experience... the existence of my agents may make my children not cherish me when I am alive...''} 

\subsubsection{Grief and Regret Endured by Family.} \label{Grief and Regret Endured by Family}

\ly{The grief and regret imagined by our participants after death emerged as a key factor influencing their attitudes toward ``AI Afterlife'' as digital legacy. }
\ly{Consistent with \cite{xygkou2023conversation} where the bereaved used chatbots of the deceased for seeking support, some desired to leverage their interactive and high-fidelity agents to alleviate the grief of their loved ones (P2-3, 6, 12, 14).}
\ly{In particular, those who experienced regrets, such as accidents, separation, or misunderstandings, during their bereavement wished that ``AI Afterlife'' could prepare their loved ones for facing similar regrets (P2-4, 11, 15-16).}
For instance, P2 regretted not preserving information about her father, who died in an accident when she was young, due to limited technology. She said, \textit{``My memories of him have faded... If I could use AI to preserve my digital being, future generations would have fewer regrets when they think of me.''}

\ly{Further, participants with more nuanced observations of grief noted that their attitudes varied with specific circumstances, such as the cause of death (P2), age at death (P4), or duration of bereavement (P9)}
They found ``AI Afterlife'' based digital legacy more necessary in cases of sudden or early death, due to the potential for greater family regret and loss, but saw it as less needed after a long, peaceful life.
P4 took her family for example, stating, \textit{``My parents are quite old and content with their lives, so there's little regret... But if I die suddenly, my husband and children would definitely need my agent to cope with loneliness.''} 
However, although \cite{xygkou2023conversation} highlights the potential of it to ease sharp grief, P9 cautioned, \textit{``My family may not accept my agent right after my death. In the future, occasionally recalling me through my agent might help them better cope with grief.''}


\subsubsection{Sustained Values for the Living in Afterlife.}

Some participants with positive attitudes envisioned ``AI Afterlife'' based digital legacy as a means to provide sustained value for their family, extending beyond grief support. Practical uses included passing on knowledge, such as cooking techniques (P1), work lessons (P13), and offering assistance, such as resolving disputes (P14) or caring for grandchildren (P3). The potential daily companionship and emotional support were also highlighted, allowing families to feel their presence and continue receiving love (P6-7, 10, 13).

\ly{However, these values were seen as limited by several factors. Some participants questioned the everyday relevance of such agents, noting that their utility might diminish over time due to outdated knowledge (P8, 15, 17) or the family's limited emotional capacity to engage with them amid busy lives (P5, 8).}
\ly{Some also worried that maintaining regular interactions with these agents might impose a mental burden and even social pressure, disrupting daily life rather than providing comfort (P4, 8, 10, 12, 17-18). }
For example, P8 explained, \textit{``As an agent represents whom family misses, they might feel guilty and conflicted if they don't use it for a long time.''} 
Further, when ``AI Afterlife'' becomes mainstream, P8 added, \textit{``Others might judge one's level of respect and nostalgia based on how often they interact with the agent''} . 


\subsubsection{Bridging Family Heritage Across Generations.}

\ly{Preserving and passing on family heritage emerged as a key motivation for creating ``AI Afterlife'' based digital legacy. Participants highlighted that AI-generated agents could play an essential role in connecting generations by maintaining and sharing cherished family memories that might otherwise fade over time (P5, 14, 16, 18).} As P16 explained, \textit{``After a person dies, they are gradually forgotten by future generations; however, AI-generated agents can preserve precious family memories, providing an interactive and evolving way to pass down family history for generations.''}

\ly{While this vision was encouraging, participants expressed hesitation due to concerns about its acceptance by younger generations, influenced by family dynamics (e.g., wishes, relationships) and social trends. For example, some were unsure whether their children would value such legacies or consider them overly sentimental (P7-8, 15-16, 18). In particular, many older participants noted a decline in younger generations' interest in spiritual family connections, attributing this shift to changing societal values (P15-18).} P17 highlighted challenges in evolving cultural contexts, \textit{``Traditional practices like funeral rituals and visiting graves during Qingming Festival\footnote{During Qingming, Chinese families visit the tombs of their ancestors to clean the gravesites and make ritual offerings to their ancestors.} are fading, as many young people prioritize advocated modern simplicity over spiritual remembrance.''}

\subsubsection{Technical Capacity, Acceptance, and Needs.}

Technical issues, such as capacity, personal openness to new technology, and individual needs, influenced participants' attitudes toward ``AI Afterlife'' based digital legacy. 
\ly{Above all, simulation quality was deemed critical, especially when compared to interactions with emotionally thoughtful real people.} P7 emphasized, \textit{``If AI can fully simulate my emotions and intelligence, I support it providing emotional comfort to my family; however, if the simulation ability is only half-baked, it could be counterproductive and hurt their feelings.''} Others similarly noted that inadequate imitation might fail to deliver positive experiences (P1, 8, 17).

\ly{Traditional attitudes toward advanced technologies also influenced participants' acceptance. Some were highly open to embracing innovative technologies (P11, 14, 16-18), even envisioning various possibilities for how ``AI Afterlife'' might shape their lives. (P11, 17-18).} For instance, P11 shared, \textit{``People make choices throughout lives. When growing old, they often find that lives are full of regrets. Therefore, I want my agent to revisit my past life moments, exploring the outcomes of different choices - what would happen if I chose to go to graduate school? or if I chose another girl as my partner?''}

\ly{Moreover, participants' attitudes were shaped by how well technology features aligned with their personal needs for digital legacy.} Some felt that remembrance through spirits, artifacts, or audio and video files was sufficient (P7-8, 17). Others believed that an AI-generated agent could better organize and preserve scattered information (P1-2, 14). As P2 explained, \textit{``It is meaningful to systematize fragmented information for preserving myself and for the nostalgia of the living.''}

\subsection{Comparing ``AI Afterlife'' with Traditional Digital Legacy}\label{comparision ai and traditional}

AI-generated and traditional digital legacy exhibit three key differences: (1) data distribution - integrated versus scattered, (2) content authenticity - generative versus authentic, and (3) interaction form - interactive versus static.
\ly{These differences underscore their unique strengths and limitations. Participants consistently emphasized that neither approach can fully replace the other due to their distinct expressive tendencies (P1, 7, 9, 12, 16). 
They advocated for a context-dependent approach, allowing users to select the form that best meets their needs or combine both approaches to maximize their complementary benefits.}

\subsubsection{Integrated and Scattered.}

AI-generated digital legacy presents an integrated approach by consolidating vast amounts of dispersed, easy-to-lose digital information, often incorporating elements of traditional digital legacy as part of AI model training. This integration enhances accessibility and usability (P2, 14), enabling users to derive more value from their digital archives, and potentially motivate the collection and preservation of traditional digital legacy.

\ly{However, limitations arise due to the exclusion of sensitive or private information}, such as passwords, which are typically shared directly and remain outside the scope of AI models (P7, 13). 
\ly{Additionally, the significant costs associated with data collection, model training, and maintaining AI-based services can pose challenges to accessibility, particularly for users with limited resources, thereby exacerbating technical inequities (P6, 16).}
As P6 observed, \textit{``For ordinary people, traditional digital legacy is more cost-effective''}.
\ly{This comparison highlights the tension between the potential benefits of integrated features of generative AI and the practical limitations related to privacy concerns and accessibility.}

\subsubsection{Generative and Authentic.}

\ly{Generative AI advancements enable agents to simulate human-like interactions with increasing realism, offering participants a sense of ongoing presence and control over their posthumous connections (P2-3, 10-11, 16).
For instance, P3 highlights the opportunities for continuity and emotional support,} \textit{``Through my AI agent, I would never truly leave the world, as I could continue to actively interact with and accompany my family.''} 

\ly{However, participants emphasized that these simulations, no matter how sophisticated, lack authenticity because they do not originate from genuine experiences or personal histories unique to the deceased  (P2, 4, 6, 14, 18). This gap in authenticity makes traditional digital legacy irreplaceable in preserving deeply personal and emotional connections. }
As P4 noted, \textit{``Traditional digital legacy like old photos that capture genuine details and history, serve as irreplaceable carriers of spiritual sustenance and cherished memories.''}
\ly{This comparison highlights the tension between the potential of realistic generative AI simulations and the irreplaceable value of traditional legacies in preserving authentic emotional connections tied to real-life experiences.
}

\subsubsection{Interactive and Static.} \label{Interactive and Static}

AI-generated digital legacy introduces interactive features that transform memories into dynamic and responsive experiences, fostering companionship and emotional support (P4, 8, 10, 13). 
P4 remarked, \textit{``This interactivity enables meaningful responses to thoughts and emotions, easing loneliness and offering comfort during grief.''} 
\ly{These interactions mimic real conversations, offering users a unique sense of connection.}

\ly{However, participants raised concerns about the potential risks of prolonged engagement, which could deepen grief instead of alleviating it. }
P8 highlighted, \textit{``Engaging with an AI agent involves back-and-forth exchanges like real conversations, entering and exiting such interactions is gradual. If abruptly interrupted, it can feel awkward and disjointed.''} 
In contrast, traditional digital legacies are static but offer the living the freedom to pause or revisit memories at their own pace, unburdened by the pressures of interaction. As P8 and P18 suggested, these static artifacts allow people to engage with memories selectively, integrating them seamlessly into daily life without emotional overwhelm.
\ly{This comparison emphasizes the need to balance the immersive interactivity of generative AI with the emotional neutrality and flexibility of static digital legacies.}

\subsection{Life Cycle of ``AI Afterlife'': Encode, Access, and Dispossess} \label{life cycle section}
\begin{figure*}[htbp]
	\centering 
	\includegraphics[width=\linewidth]{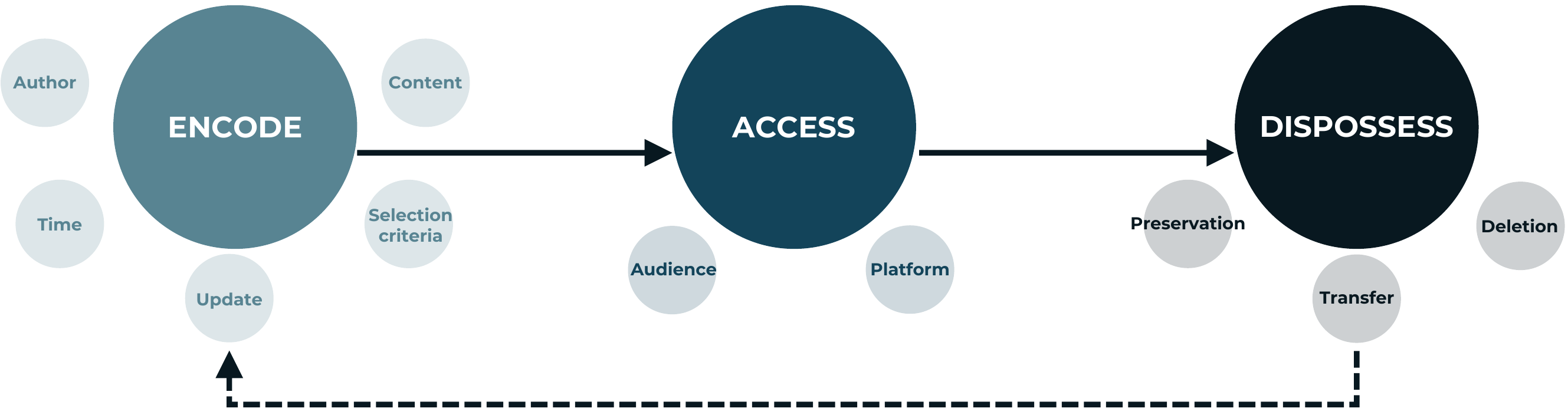}
	\caption{Life cycle of ``AI Afterlife'' as digital legacy, including encoding the legacy content into a digital form, accessing that content, and the dispossession of that legacy to another individual or through deletion.}
	\label{fig:life cycle}
        \Description{}
\end{figure*}

As proposed in \cite{doyle2023digital}, the life cycle of digital legacy consists of three stages: encode, access, and dispossess. ``Encode'' refers to converting legacy content into a digital format, ``access'' involves retrieving that content, and ``dispossess'' pertains to its transfer or deletion. The themes identified in our study align with these stages, allowing us to categorize our findings accordingly. This alignment is illustrated in \autoref{fig:life cycle} and summarized in \autoref{tab:summary of themes and main points in life cycle}.

\begin{table*}[htbp]
  \caption{Summary of themes and main points in the life cycle of ``AI Afterlife'' as digital legacy.}
  \label{tab:summary of themes and main points in life cycle}
  \begin{tabular}{p{0.1\textwidth}p{0.85\textwidth}}
    \toprule
    \textbf{Theme} & \textbf{Main Point} \\
    \midrule
    \textbf{Encode}&\\
    \hline
    \multirow{3}{=}{Author} 
            & Manage the creation themselves for security, simulation accuracy, preferences (P1,3,6-7,10-14,16-18) \\ 
             &\cellcolor{gray!15}Allow family to retain the creation right in case age-related declines or unforeseen accidents (P2-4,7,14)  \\ 
             & Leave the decision to family, allowing flexibility in creation (P8,12,18) \\
    \hline
    \multirow{4}{=}{Time} 
            & \cellcolor{gray!15}Create in daily life to get enough data preparation time and better control over agents (P2,4,7,10-11,17) \\
            & Create in later life or near death for necessity (P1,15) \\ 
             & \cellcolor{gray!15}Leave it posthumously due to concern about family's acceptance and needs (P8,12,14,18) \\ 
             & Postpone the process out of prioritizing family needs during life (P4,6) \\
    \hline
    \multirow{5}{=}{Content}
            & \cellcolor{gray!15}Extensive existing digital information (P1-18) \\ 
            & Document personal details such as physical attributes and habitual behaviors (P1, 3, 6, 10) \\ 
            & \cellcolor{gray!15}Incorporate data from close family (P1-2, 4, 6, 8, 14) and social connections (P7, 10, 13, 16, 18)  \\ 
            & Insufficient data for simulating deeper human attributes, expect for P17 \\ 
            & \cellcolor{gray!15}Challenging for those willing to record and organize deeper information (P3-4,6-7,11,13,17) \\
    \hline
    \multirow{5}{=}{Selection criteria}
            & Sincerity is essential to ensure authenticity (P1,7-8,18) \\
            & \cellcolor{gray!15}Daily trivia is viewed as irrelevant for future use (P13,17) \\ 
            & Disgraceful experiences are commonly excluded (P13-14,16) \\ 
            & \cellcolor{gray!15}Painful memories and imperfect personality traits are filtered out to avoid unpleasantness (P4,6,13) \\ 
            & Confidential information, such as bank passwords, was universally excluded (P1-18) \\ 
    \hline
    \multirow{4}{=}{Update}
            & \cellcolor{gray!15}Incorporate new data or address inconsistencies arising from real-life interactions (P1-18) \\ 
            & Keep the update right private or limited to their closest family (P10-11,14,16) \\ 
            & \cellcolor{gray!15}Extended the update right to a wider circle of relationships (P7,11-13,16,18) \\ 
            & Updates often undergo review by an assigned manager with family-centered priorities (P7,13) \\
    \hline
    \textbf{Access}&\\
    \hline
     \multirow{3}{=}{Audience} 
            & \cellcolor{gray!15}Open-minded about who has access, prioritizing the wishes of others (P1,3,8,15,18). \\ 
             & Impose restrictions by closeness (P10,12), communication value (P2,4,6,13,16), and risks (P6,9,13-14) \\ 
             & \cellcolor{gray!15}Different views in who can set access permissions - self (P1-18), maintainer (P7,13,16), AI (P1,7) \\
    \hline
    \multirow{3}{=}{Platform} 
            & Virtual spaces, like social media, memorial websites, mobile apps, XR apps (P1-18) \\ 
             & \cellcolor{gray!15}Robots in the domestic setting, controlled by an assigned manager (P6,9,16) \\ 
             & Combined with traditional memorial artifacts, like gravestones (P1) and plaques in ancestral halls (P6) \\
    \hline
    \textbf{Dispossess}&\\
    \hline
    \multirow{3}{=}{Preservation or deletion} 
             & \cellcolor{gray!15}Based on the wishes of the living (P2,6-10,16-17) \\ 
             & Based on the depth of communal connections and affection (P3-4,14,17-18) \\ 
             & \cellcolor{gray!15}Preserve permanently with concerns about reputation, capabilities, and maintenance cost (P1-2,9)\\
    \hline
    \multirow{3}{=}{Transfer} 
            & Specifying transfer details in the will (P10-11)  \\ 
            & \cellcolor{gray!15} Rights transferred, such as author (P6-7, 10-12), update (P10-11, 14), and audience (P1-2, 7) \\
            & Closest family members are the only recipient choice (P1-18) \\ 
    \bottomrule
  \end{tabular}
\end{table*}

\subsubsection{Encode}

In the context of ``AI Afterlife'' as digital legacy, ``encoding'' refers to the process of gathering and using digital information to train generative AI models that simulate realistic human beings. Key aspects of the encoding process include author, time, content, selection criteria, and updates.

\textbf{Author.} \label{author}
\ly{Many participants preferred to manage the creation process rather than leaving the whole task to their families.} They believed self-creation ensured data security (P1, 3, 6), maintained simulation accuracy (P6-7, 10, 12), and aligned with personal preferences (P13, 14, 16, 18). P6 emphasized the importance of controlling data to prevent privacy leaks, while P7 noted, \textit{``Without my involvement, the simulation might be biased and lack authenticity.''}
\ly{However, in anticipation of potential challenges such as age-related decline (e.g., mobility issues, decreasing digital skills) (P3-4, 14) or unforeseen accidents (P2, 7), they considered allowing their trusted family to retain the creation right, ensuring that the creation would align with their intentions even in the absence of their active involvement.}
P2 mentioned, \textit{``It doesn't matter who created my agent because it's impossible to predict whether tomorrow or an accident will come first.''} She believed her role was to keep her digital information organized for future flexibility.
\ly{Additionally, some who were less motivated to create agents for themselves and primarily for family purposes, preferred to leave the decision to their family, allowing them to create the agents as they wished. Uncertainty about whether their family would accept the agent further reinforced this choice (P8, 12, 18).} P12 explained, \textit{``It's not something obligatory... If my family wants it, they can create it themselves. If they don't feel connected to me, making it for them would be meaningless.''}

\textbf{Time.}
\ly{Many participants preferred involving the creation process as part of daily life (P2, 4, 7, 10-11, 17), rather than waiting until later life or near death (P1, 15).} 
This aligns with findings under the ``Author'', where earlier involvement allowed for better control over their agents. 
Besides, P7 noted that prolonged data collection led to more realistic simulations. Similarly, P11 emphasized, \textit{``It is necessary to collect data 10 years in advance... It is impossible to collect data one or two months before death.''} 
\ly{In contrast, some considered delaying the creation after death, leaving the creation decision entirely to their family due to the concerns about their acceptance and needs (P8, 12, 14, 18), or prioritizing family needs over self agent creation during life (P4, 6).}
For example, P6 expected to postpone creating her agent until completing her parents' agents, while P4 prioritized spending time with family and expressed concern, \textit{``My agent might disrupt my family's normal life after my death... It should be created only after my family has recovered from grief.''}

\textbf{Content.} \label{content}
Most participants could provide extensive data stored on digital devices, such as media (photos, audio, videos), files (writing samples), and social media accounts (chat history, updates). Many were also willing to document personal details such as physical attributes (e.g., height, weight, skin texture) and habitual behaviors (P1, 3, 6, 10). Despite challenges in data recording and preservation, such as insufficient attention (P8, 16), data loss during migration, and deletion due to limited storage (P1, 5, 15), the available data is generally sufficient for simulating users' physical features and facilitating basic conversations, owing to advancements in generative AI. Most participants were also open to incorporating data from close family (P1-2, 4, 6, 8, 14) and social connections (P7, 10, 13, 16, 18) to enhance simulation performance.
However, simulating deeper human attributes, such as personal experiences and inner thoughts, presents greater challenges. Only P17, who has maintained a daily diary since his twenties, could provide such data directly. While others expressed willingness to record details of their experiences and thoughts (P3-4, 6-7, 11, 13, 17), barriers such as memory gaps and the time required for organization were significant. 
For example, P18 was open to providing existing data but hesitant to organize her life story, stating, \textit{``I don't like writing down my thoughts or experiences. Daily communication already covers them. Creating my AI agent doesn't motivate me to spend time on this.''}

\textbf{Selection criteria. } \label{selection criteria}
\ly{Although the importance of sincerity to ensure their agents remain authentic is emphasized (P1, 7-8, 18), most participants considered several factors when selecting digital information for AI-generated agents, including daily trivia, disgraceful experiences, painful memories, imperfect traits, and confidential data.} Many deemed daily trivia irrelevant for future use (P13, 17), with P17 stating, \textit{``People interacting with my agent would likely be more interested in my thoughts than in what I did at a specific time.''} \ly{Disgraceful experiences, such as sharp remarks, conflicts of interest, and harm caused to others, were expected to be excluded to avoid potential ethical, economic, or legal issues (P13-14, 16).} Painful memories (P4, 13) and imperfect traits (P6) were also filtered out to minimize unpleasant interactions. P6 explained, \textit{``I want my agent to get along better with my family because I have a bad temper... I don't want to bring back the unpleasant memories of our quarrels.''} Confidential information, such as bank passwords, was universally excluded, as P7 stated, \textit{``They are just secrets that have no meaning in representing my true self.''} 

\textbf{Update.} \label{update}
\ly{Most participants emphasized the necessity of updating their agents after the initial version was created, either to incorporate new data or address inconsistencies arising from real-life interactions to enhance the agents' comprehensiveness and authenticity.}
\ly{Opinions on update permissions were diverse, reflecting a balance between individual control and relational dynamics. }
Some participants preferred to limit this right to themselves or their closest family members (P10-11, 14, 16). For example, P14 stated, \textit{``I don't want others to update my agent, as everyone has different ideas and may not understand me; otherwise, it might not align with my intentions.''} P16 extended this right to his son, who knew him well.
In contrast, other participants extended the update right to a wider circle of relationships (P7,11-13,16,18), with P7 noting, \textit{``A person is defined by his social connections.''} 
These updates were often subject to review by a trusted manager (P7,13). 
\ly{During the review process, family-centered priorities often took precedence over strict authenticity to ensure the agent served the emotional and practical needs of the living.}
For instance, P13 explained, \textit{``The purpose of creating my agent is to accompany my son, so the updates must be acceptable to him. If he finds them distasteful, they will not be approved.''}

\subsubsection{Access}
In the context of ``AI Afterlife'' as digital legacy, ``access'' refers to who can access and how to access AI-generated agent-based digital legacy. We identify two key aspects - audience and platform in this process. 

\textbf{Audience.} \label{audience}
\ly{Participants indicated that access to their agents is typically granted in descending order of priority: close family, common connections, and strangers. There are several factors influencing access permissions significantly.} 
\ly{Some were open-minded about who could access their agents, prioritizing the wishes of others (P1, 3, 8, 15, 18).} For instance, P8 remarked, \textit{``I'm flexible about who has access rights. Only those with a deep relationship with me will remember and miss me. If the relationship is ordinary, they likely wouldn't interact with the agent of a deceased individual.''}
\ly{In contrast, those prioritizing practical values and risk prevention set restrictions based on relationship closeness (P10, 12), communication value (P2, 4, 6, 13, 16), and concerns about reputation, security, and legal issues (P6, 9, 13-14).} P6 restricted access for strangers, explaining, \textit{``Few strangers would be interested in my story... there is also a risk of data leakage.''} P9 highlighted legal concerns, stating, \textit{``Currently, the legal rights and obligations of this technology are undefined... \ly{For example, if it infringes upon others' rights in chat, determining responsibility can be difficult... From a risk management perspective, access should be restricted.''}}
\ly{There were also differing views on who should set access permissions.} All agreed that the individual being represented should initially hold this right. Once the legacy is transferred, those who maintain it should manage it (P7, 13, 16). However, some suggested once AI-generated agents develop their own thoughts, they should have priority (P1, 7). P1 stated, \textit{``I hope my agent can choose who to interact with and for how long, rather than simply following fixed rules... It should behave like an AI agent with its own personality, not just an AI assistant.''}
\ly{It provides a glimpse into a more complex issue of autonomy after AI agents have developed independent consciousness.}

\textbf{Platform.}
\ly{Participants envisioned a diverse set of digital platforms for their agents to be accessed.}
\ly{Most participants supported their agents to be accessed in virtual spaces, through various commonly, easy-to-use online platforms such as social media, memorial websites, and specific mobile or Mixed Reality (XR) applications.} Among them, WeChat \cite{WeChat2024}, one of China's most popular social media, was been regarded as the most suitable platform to embed it due to convenience and familiarity (P1, 3-4, 7, 10, 13-14, 16). P16 explained, \textit{``Since most people communicate through WeChat, it would be very convenient if my family and friends could access my agent directly through it...They could also use familiar chat and audio/video call functions to interact with my agent.''} P14 added, \textit{``Using other platforms complicates things (e.g., requiring new accounts). In contrast, WeChat is great for remembrance, as it retains previous chat history.''} 
Besides, P11 preferred embedding his agents in XR glasses, while others had no particular preference in these virtual spaces.
\ly{Another significant consideration was the use of robots to access these agents at home, managed by an assigned person (P6, 9, 16), offering a tangible presence for interactions (more details in section \ref{embodiment})}
\ly{Some participants also proposed combining AI agents with traditional memorial artifacts, such as gravestones (P1) and plaques in ancestral halls (P6). This approach preserves traditional commemorations while recreating their meaning in new contexts.}
P6 envisioned, \textit{``If ancestral halls had plaques with AI-generated agents of older generations, it could create a sense of large family during gatherings... It could also preserve and pass down family heritage in a more personal and face-to-face way.''}

\subsubsection{Dispossess}

In ``AI Afterlife'' as digital legacy, ``dispossess'' refers to transferring the AI-generated agent-based digital legacy, or deleting them when no longer needed. We identify deletion conditions and transfer preferences.

\textbf{Preservation and deletion.}
\ly{Participants consider whether to preserve or delete the agents based on the living's wishes (P2, 6-10, 16-17) and the depth of their relationships (P3-4, 14, 17-18).} P16 speculated that three generations might be the maximum for transferring his agents, stating, \textit{``I respect users' wishes when deleting it... Once there's no willingness to use it, its meaning fades.''} P3 added, \textit{``Only when people are close will they develop a desire to communicate.''} 
\ly{In contrast, some wished to preserve their agents permanently but were concerned about reputation, capabilities, and maintenance costs (P1-2, 9).} P2 worried, \textit{``If I were a person of prestige in my family or society, I might pass down it through generations... but I'm just an ordinary person without the means to maintain it.''} 

\textbf{Transfer.}
\ly{Transferring the agents with essential details between generations is significant, such as the approaches, recipients, and rights.} P10 and P11 emphasized the importance of specifying transfer details in their wills, stating, \textit{``The transfer process should respect my wishes. I would set them in advance, such as who can inherit them and what rights can be transferred.''} 
Most participants preferred to pass on their agents to close family members. In terms of rights transferred, they typically include topics mentioned above in section \ref{life cycle section}, such as author assignment (P6-7, 10-12), update rights (P10-11, 14), and audience permission (P1-2, 7).

\subsection{Interaction Design: Replicating, Simulating, and Responding}
\label{sec: interaction design}
Exploring the interactive nature of ``AI Afterlife'' as digital legacy, we delved into how it replicates, simulates, and responds to aspects of self and interactions with external environments. We identified six themes as shown in \autoref{tab:summary of themes and main points in interaction}: age and appearance, embodiment, simulation level, expression of extra knowledge, evolution, and reactivity.

\begin{table*}[htbp]
  \caption{Summary of themes identified in participants' anticipated interaction designs for their ``AI Afterlives''.}
  \label{tab:summary of themes and main points in interaction}
  \begin{tabular}{p{0.15\textwidth}p{0.8\textwidth}}
    \toprule
    \textbf{Theme} & \textbf{Main Point} \\
    \midrule
    \multirow{5}{=}{Age and Appearance} 
            & Customized to reflect the most familiar and meaningful stages for the intended audience (P2,3,6) \\ 
             & \cellcolor{gray!15}Customized to be status shortly before death to continue their presence (P5,7)  \\ 
             & Customized to be healthy and energetic status to avoid evoking unsettling memories (P3,6,16) \\ 
             & \cellcolor{gray!15}Customized to be other expected status like youthful, mature, and pleasing (P2,4,8,10,13-14,16) \\ 
             & Dynamically adaptive to different users and interaction needs (P1-2,6,11-12,14-18) \\
    \hline
    \multirow{6}{=}{Embodiment} 
            & \cellcolor{gray!15}Virtual forms offer more controllable engagement with memories of loved ones (P3-4,8) \\ 
            & Virtual forms are sufficient for remembrance (P6,18) \\ 
             & \cellcolor{gray!15}Physical forms associated discomfort (P4,6,9)  \\ 
             & Physical forms related ethical concerns, particularly how to treat AI-generated agents (P7,18) \\ 
             & \cellcolor{gray!15}Physical forms also have a higher maintenance cost (P6) \\
             & Adaptability of forms based on age, individual needs, and the duration of grief (P6,9,14) \\
    \hline
    \multirow{3}{=}{Simulation Level} 
            & \cellcolor{gray!15}Simulating the surface features is not enough (P1-5,7-18) \\ 
             & It is needed to enable AI-generated agents with thinking ability (P1-8,10-18) \\ 
             & \cellcolor{gray!15}Advocate for a balance between authenticity and discretion (P1,7-8,18) \\
    \hline
    \multirow{3}{=}{Expression of Extra AI Knowledge} 
            & Create discrepancies from real self in aspects like knowledge, capabilities and values (P1,16-17) \\ 
             & \cellcolor{gray!15}Enhance usefulness and interaction quality by adding knowledge and cultural depth (P6,12,18)  \\ 
             & Balance between personal characteristics and additional knowledge (P8,11,13-14) \\
    \hline
    \multirow{3}{=}{Evolution} 
            & \cellcolor{gray!15}Stay presence and maintain relevance in communication (P1,3-4,6-8,12-14,16-18) \\ 
             & Bring uncontrollable risks, authenticity and identity issues, and philosophical reflection (P9-11) \\ 
             & \cellcolor{gray!15}Adaptive based on life stages (P2-4) \\
    \hline
    \multirow{3}{=}{Reactivity} 
            & Autonomously sense and engage with their environment (P1-3,6,11) \\ 
             & \cellcolor{gray!15}Potential intrusiveness of active reactivity (P4,9-10,12-14,16-18) \\ 
             & Advocating for a balanced approach based on needs, relationship, and preparation (P4,7-8,11) \\
    \bottomrule
  \end{tabular}
\end{table*}

\subsubsection{Age and Appearance} \label{age and appearance}
Most participants wanted to customize their agents' age and appearance to reflect familiar and meaningful life stages. Many preferred their agents to represent key moments for their intended audience (P2, 3, 6). For example, P2 said, \textit{``If my agent is for my children, I would choose the time when I took care of them... the most profound memory is when they were with us before college.''} Some also preferred to depict themselves shortly before death to maintain a sense of presence (P5,7), as P5 explained, \textit{``It feels like I haven't left them and can continue growing with them.''} Others wanted a healthy, energetic appearance to avoid negative associations with declining health (P3, 6, 16), or considered youthful, mature, or pleasing looks (P2, 4, 8, 10, 13-14, 16).

In contrast, some participants suggested that their agents should dynamically adapt based on personalized experiences with different users (P1-2, 6, 11-12, 14-18). P12 explained, \textit{``As I interact with different people at various life stages, the memories I share with each occur at different times, so I can't present the same image to everyone.''}

\subsubsection{Embodiment}\label{embodiment}

Most participants preferred virtual forms over physical embodiment for comfortable remembrance, ethical concerns, and affordability. Virtual agents allowed controllable engagement with memories (P3-4, 8). P3 explained, \textit{``With a virtual agent, my family can reach out when they miss me... I don't want a physical presence constantly reminding them of my death.''} Many felt virtual agents were sufficient, with P18 noting, \textit{``The connection between my family and me is primarily spiritual, not physical.''}
Participants also expressed discomfort with physical forms (P4, 6, 9). P4 stated, \textit{``When seeking comfort, there's an awareness that a physical agent isn't truly the loved one, which can induce fear.''} Ethical concerns included managing the physical agents of the deceased (P7, 18). P7 highlighted complexities in \textit{``handling and storing them resembling family members.''} P18 added, \textit{``Physical forms might show disrespect for the body... breakage could cause trauma.''} Economic factors were also noted, with P6 mentioning higher maintenance costs.

\ly{Adaptability based on age, needs, and grief duration was also discussed (P6, 9, 14). P9 noted that her physical agent could better support her husband in old age due to accessibility and lower digital literacy. Similarly, P6 noted, \textit{``Physical forms might support my young children but become burdensome over time.''} P14 added that while physical forms offer temporary comfort, they may not sustain long-term emotional well-being.}

\subsubsection{Simulation Level} \label{simulation level}

Different simulation levels during interaction have been explored, including external physical features, thinking, and privacy. Few participants believed simulating physical features alone is sufficient, except for P6, who suggested, \textit{``Replacing the voice of a voice assistant with my own, such as in navigation apps, would make interactions feel more personal and engaging... ChatGPT's voice always feels cold and lacks variety''}

\ly{Besides, participants agreed that agents should exhibit personal ``thinking'' ability. P8 stated, \textit{My agent should reflect my personal characteristics and family knowledge, generating content based on my style, beliefs, and values. If it only has AI knowledge, interacting with it would be no different from using ChatGPT, which is confusing.''} P9, with a background in philosophy, noted, \textit{``Philosophically, life is defined by thought, so achieving this simulation level may require rethinking boundaries of life and death, along with related social, legal, and societal issues.''}}

Regarding privacy, participants advocated for a balance between authenticity and discretion (P1, 7-8, 18). While some preferred minimal data filtering for a more authentic simulation rather than a ``saint simulation'', they cautioned against including much sensitive information to protect privacy. 

\subsubsection{Expression of Extra AI Knowledge} \label{ai knowledge}

Participants had mixed feelings about incorporating external knowledge embedded in large foundational models into their agents. Concerns focused on its impact on interaction quality and authenticity (P1, 16-17). P1 noted, \textit{``Embedding this knowledge may lead to a misrepresentation and incomplete understanding of values and knowledge...''} P16 shared a similar view, stating additional knowledge might create discrepancies between the agent and the real self, but also noted, \textit{``Not including it may make the agent seem outdated.''} P17 questioned the practicality, saying, \textit{``It might discuss topics with my children that I'm unfamiliar with, making interactions feel odd.''}

On the other hand, integrating AI knowledge could enhance interaction quality (P6, 12, 18). P6 believed it would make the agent more useful and appealing to her family, while P18 felt it \textit{``enriches the agent's knowledge and cultural depth.''} This was particularly useful when \textit{``interaction scenarios require feedback that can't be predetermined''} (P12).

Some participants favored a balance between personal traits and added knowledge (P8, 11, 13-14). P8 suggested blending personal traits with extra information, rather than relying solely on AI knowledge. Similarly, P11 emphasized, \textit{``Extra AI knowledge should align with my perception of knowledge.''} P14 welcomed the integration, saying, \textit{``As long as it enhances interaction quality without compromising my essence.''} P13 also noted the need to avoid negative traits, adding, \textit{``If the AI knowledge is more positive and insightful than mine, it could be beneficial.''} 

\subsubsection{Evolution} \label{evolution}

Participants emphasized the need for a balance between authenticity and adaptability in agent evolution (e.g., age and thinking) over time. Many supported the idea of allowing agents to evolve to stay present and maintain relevance in communication (P1, 3-4, 6-8, 12-14, 16-18) as evolving agents could better adapt to new contexts. \ly{P8 stated, \textit{``My agents should understand my family updates and the evolving world around us, providing relevant interactions.''} P4 echoed, \textit{``It would be ideal for my agent to continue meaningful conversations, as if I were still present.''}}

In contrast, some raised concerns about the risks and philosophical implications, particularly regarding authenticity and identity (P9-P11). P10 preferred static agents, saying, \textit{``It should remain as designed to preserve its essence, avoiding new topics that might seem inauthentic.''} P11 echoed these concerns, noting that evolution could introduce uncontrollable changes. P9 questioned the philosophical implications, stating, \textit{``Evolving agents challenge individual identity consistency... It's important to consider what makes someone the same person despite physical, psychological, or digital changes.''}

Some also emphasized adapting based on life stages (P2-4). P4 noted, \textit{``Further aging for the agents of the elderly is inappropriate.''} While P2 added the younger might benefit from continued growth.

\subsubsection{Reactivity} \label{reactivity}
\ly{
Reactive interaction highlights the balance between autonomy and non-intrusiveness.
Some participants expressed interest in active reactivity, suggesting it as a feature motivated by love for their family (P1-3, 6, 11).} 
For instance, P2 noted it would be comforting if the agent detected family emotions and provided support. P3 added, \textit{``My family wouldn't mind having sensors at home for my agent to respond to their needs in real-time.''}

\ly{Conversely, many preferred passive reactivity (P4, 9-10, 12-14, 16-18), emphasizing the importance of respecting personal space and avoiding disturbing the normal life of their families.} P16 stated, \textit{``It would be unsettling if my agent \ly{(robot form)} was walking around the house without being explicitly called upon...''}

\ly{Confronted with the contradiction of non-intrusiveness and autonomy, some suggested a balanced approach (P4, 7-8, 11), proposing that agents remain largely passive while providing comfort when needed by the family.}
P8 noted that reactivity might depend on the interaction context and the relationships between the deceased and their loved ones. P7 raised ethical concerns, emphasizing the need for family involvement in customizing this function to avoid unintended consequences. 
\ly{Those interested in active reactivity also saw it as an optional choice, with P11 explaining, \textit{``My family should have the right to customize this function.''}}

\subsection{Concerns: Technology, Mental Health, Security, and Socioeconomic Issues}\label{concerns}

Several significant concerns regarding ``AI Afterlife'' as digital legacy have been identified, including technology, mental health, security, and socioeconomic issues, as summarized in \autoref{tab:summary of themes and main points in concerns}.

\begin{table*}[htbp]
  \caption{Summary of themes and main points in concerns.}
  \label{tab:summary of themes and main points in concerns}
  \begin{tabular}{p{0.15\textwidth}p{0.8\textwidth}}
    \toprule
    \textbf{Theme} & \textbf{Main Point} \\
    \midrule
    \multirow{3}{=}{Technology} 
             & Technological limitations of AI in simulating realistic human (P1,7,11-12,17-18)  \\ 
             & \cellcolor{gray!15}{Unpredictability and potential loss of control over AI systems (P9-11,13)} \\ 
             & Lack of standards and the challenges in regulating the proper use (P7,P12)  \\ 
    \hline
    \multirow{3}{=}{Mental Health}
            & \cellcolor{gray!15}{The blending of real and virtual worlds could evoke feelings of fear and discomfort (P5,7-8,16)} \\ 
            & Interactions with the digital being of the deceased could hinder the healing process (P2-5,8,10,12,16,18) \\ 
            & \cellcolor{gray!15}{Design technology features that encourage moderation in interactions and move on  (P2,12,16,18)} \\
    \hline
    \multirow{3}{=}{Security}
            & Handling of personal data and privacy in the creation and access stages (P2-4,6,11,13) \\ 
            & \cellcolor{gray!15}{AI-generated agents negatively impact reputation (P11, 13-14, 16-17)}  \\ 
            & Abused by malicious actors for harmful purposes, like fraud (P10-12,17-18) \\ 
            & \cellcolor{gray!15}{Abused by service providers without proper oversight (P1,2)} \\
    \hline
    \multirow{3}{=}{Socioeconomic Issues}
            & Impact of AI-generated agents on social interactions (P5,8,10) \\ 
            & \cellcolor{gray!15}{Impact traditional cultural practices surrounding death and mourning (P6,9,16)} \\ 
            & Economic implications of creating and maintaining AI-generated agents (P5,11,15-16) \\
    \bottomrule
  \end{tabular}
\end{table*}

\subsubsection{Technology}
Technical concerns include AI's limitations in simulating human traits, the unpredictability of autonomous systems, and the lack of regulatory standards.
Many noted that incomplete or biased input data could create flawed or skewed agents (P1, 7, 11-12, 17-18). \ly{For example, P1 explained, \textit{``AI may be influenced by some default values... Besides, although its capabilities are relatively comprehensive, it may lack the depth in my area of expertise... The data I provide is also limited and cannot reflect my complete life.''}}

Unpredictability and potential loss of control were also significant worries (P9-11, 13). Participants feared that AI systems could evolve beyond their original settings, causing unforeseen ethical or social consequences. For instance, P13 mentioned, \textit{``If my agent develops harmful tendencies, it could negatively impact my son's emotions and well-being.''}

Additionally, the lack of regulatory standards raised concerns about proper use and oversight (P7, 12). P12 stated, \textit{``It's challenging to educate and regulate users on responsible usage due to freedom and accessibility of this technology.''}

\subsubsection{Mental Health}
Mental health concerns center on fear and obsession with grief.
Some noted that blending the real and virtual worlds often evoked discomfort (P5, 7-8, 16). P8 noted, \textit{``Long-term use can cause emotional turmoil... the presence of the deceased can trigger uncomfortable psychological reactions.''} The ``uncanny valley'' effect was also highlighted, which caused unsettling feelings (P7, 16), especially for those anxious or superstitious (P16).

Obsession with interacting with agents of the deceased was another concern, as it could hinder healing and moving on from grieving (P2-5, 8, 10, 12, 16, 18). P5 shared, \textit{``It's easy to become immersed in grief if interacting with these agents.''} P2 added, \textit{``This has diverged from my intent of creating my own agent.''} P8 pointed out that these agents provided a more immersive experience than photos or memories, potentially disrupting daily life.

To mitigate these risks, participants suggested features to moderate interactions (P2, 12, 16, 18). P12 recommended limiting realism, like using silent video, to reduce immersion. P16 proposed adding reminders to prevent excessive mourning, with detection features to identify when the living are overly sad. P18 emphasized his concern for his family, stating, \textit{``While leaving an agent for my family may help them remember me, I hope it encourages them to move on.''}

\subsubsection{Security}

Security concerns mainly focus on privacy, reputation, and potential abuse. A key issue was the handling of personal data in creating and accessing these agents (P2-4, 6, 11, 13). P2 emphasized, \textit{``My digital information should remain private, not made public.''} P6 added, \textit{``The interactions with my agents should not be accessible to strangers.''}

Some also feared the potential harm to their reputation (P11, 13-14, 16-17). While some tried to filter out disgraceful experiences during the creation process to prevent exposure (P13-14, 16), concerns remained, as P11 notes, \textit{``If my agent reveals my darker side, it could damage my reputation and negatively impact my family.''}

Additionally, concerns about misuse were prominent, particularly the risk of agents being exploited for fraud (P10-12, 17-18). P10 warned, \textit{``Malicious individuals could exploit my agents for profit.''} P12 highlighted the difficulty of managing misuse, stating, \textit{``Images and voices can be easily extracted from my agents.''} P17 emphasized the need for safeguards, \textit{``I lived my life doing good. It's unacceptable for my agents to engage in wrongdoing.''} \ly{Concerns about abuse also extended to service providers. Some expressed expressed distrust toward private, profit-driven companies, citing issues with oversight and sustainability (P1-2). As P1 explained, \textit{``A well-managed organization might succeed, but without proper oversight, my agent could become an electronic doll. It would be preferable if a non-profit organization, using open-source technology and free from capital influence, were responsible, with adequate government and public supervision...''}}

\subsubsection{Socioeconomic Issues}
Socioeconomic concerns involve social interactions, cultural implications, and economic factors. Participants had mixed feelings about the impact of ``AI Afterlife'' on social interactions (P5, 8, 10). P8 mentioned the guilt and social pressure the family might face if the agent is not used long-term, while P5 warned it could harm real-world interactions. P10 added, \textit{``People are already addicted to the internet, sitting together but focused on their phones. Using AI agents of the deceased may lead to a future where people no longer interact with real people, which is troubling. Living should involve real-world experiences, not just virtual ones.''}

AI-generated agents are also expected to disrupt traditional cultural practices around death and mourning (P6, 9, 16). P9 noted, \textit{``AI agents for the deceased will change how we conduct funerals and memorials.''} P6 added that it could \textit{``allow the deceased to join family reunions for remembrance.''}

Additionally, economic concerns about creating and maintaining these agents were raised (P5, 11, 15-16). P5 said she would consider creating an agent if costs were manageable, while P11 pointed out the high data requirements, \textit{``If data collection costs are too high, the agent may not be feasible.''} P15 expressed concern over ongoing maintenance costs.

%% file: sections/05-Discussion.tex
\section{Discussion}

\ly{In this section, we situate ``AI Afterlife'' in digital legacy, and delve into design implications for maintaining identity consistency and balancing intrusiveness and support in future ``AI Afterlife'' based digital legacy practices.}

\subsection{Situating ``AI Afterlife'' in Digital Legacy}

Our work focuses on ``AI Afterlife'', an emerging form of digital legacy that leverages the generative, interactive, and dynamic capabilities of generative AI. Unlike traditional static digital legacy, this new form offers transformative possibilities in two key areas: (1) maintaining a sense of continuity after death by emphasizing identity consistency (in section \ref{identity consistency}), and (2) enabling interaction with the living, particularly fostering a healthy ``continuing bond'' with loved ones by balancing intrusiveness and support (in section \ref{balance intrusiveness and support}). In alignment with the concept of ``digital immortality'' defined by Bell \textit{et al.} \cite{bell2001digital}, this new form of digital legacy achieves a degree of ``two-way immortality''. 

In addressing this emerging form of digital legacy, we first identified the diverse and complex attitudes people hold toward it, shaped by personal, familial, technological, and social factors (in section \ref{attitudes factors}). Notably, ``AI Afterlife'' challenges traditional definitions of life and death, while the interplay between these perspectives significantly shapes participants' attitudes toward this new form of digital legacy. Moreover, ``AI Afterlife'' presents opportunities to positively impact the living by supporting the grieving process, providing enduring value, and preserving family heritage across generations. However, these benefits are also contingent upon family dynamics, technological capabilities, and broader social contexts.

Besides, we examined how people perceive the differences between ``AI Afterlife'' based digital legacy and traditional digital legacy (in section \ref{comparision ai and traditional}). Three key distinctions emerged: data distribution, content authenticity, and interaction forms. This comparison highlights the tension between the two approaches. On one side are the benefits of generative AI, such as integrated information, realistic simulations, and immersive interactivity. On the other side are the advantages of static digital legacy, including accessibility, emotional connections to authentic experiences, and controllable interaction. These differences reveal their unique strengths and limitations. They suggest the need for context-dependent decisions - choosing the approach that best suits users' needs or combining both to maximize their complementary benefits.

Additionally, ``AI Afterlife'' sheds light on the design aspects throughout the life cycle (in section \ref{life cycle section}) and interaction process (in section \ref{sec: interaction design}) of digital legacy.
Aligning with the framework used for traditional digital legacy \cite{doyle2023digital}, our work further unpacks key aspects of this new form across the encoding, accessing, and dispossession processes, providing comprehensive considerations for the entire life cycle of future digital legacy practices. The unique characteristics of generative AI also introduce distinct interaction dynamics in replicating, simulating, and responding to both the self and external environments.

Furthermore, ``AI Afterlife'' introduces distinct concerns regarding digital legacy, particularly in terms of technology, mental health, security, and socioeconomics, shaped by the unique features of generative AI, as well as personal and contextual issues (in section \ref{concerns}). Therefore, careful consideration is necessary to address these potential issues and ensure that this emerging form of digital legacy remains reliable and has a positive impact on both the deceased and the living.

In summary, ``AI Afterlife'' represents a pivotal evolution in the fields of digital legacy. By exploring the complex attitudes, differences, life cycles, interaction design, and challenges, this study highlights the transformative potential of generative AI in shaping digital legacy. Building on this understanding, we propose more targeted design implications for future practice, which will be illustrated in detail in section \ref{identity consistency} and section \ref{balance intrusiveness and support}.

\subsection{Design Implications for Maintaining Identity Consistency} \label{identity consistency}

Extending life and values beyond death is a crucial factor shaping positive attitudes toward ``AI Afterlife'' as a form of digital legacy (see Section \ref{attitudes factors}). This post-mortem continuity is intrinsically linked to the concept of ``identity consistency'', which refers to the preservation of core traits, such as appearance, memory, thoughts, and beliefs, despite changes over time and across contexts \cite{wyness2019childhood, marcia1966development, taylor1989sources}. By ensuring the stability of these core traits, individuals can continue to be recognized as the ``same person'' in evolving circumstances \cite{hume2000treatise, locke1847essay, wyness2019childhood}. As ``AI Afterlife'' serves as an emerging form of digital legacy of individuals, maintaining the consistent identity of the deceased within these agents becomes a fundamental design consideration. Insights from results highlight three dimensions of identity - appearance, knowledge, thinking, and emphasize the need to manage evolution while preserving identity consistency.

\textit{Appearance.} 
While technology enables the realistic simulation of human physical traits (in section \ref{content}), how these traits are presented significantly affects identity consistency. Participants preferred to display their most meaningful or ideal appearances while allowing audiences the freedom to interpret or choose based on personal memories (in section \ref{age and appearance}). This flexibility underscores identity consistency as a dynamic process, balancing the stability of core traits with individual expression in changing contexts. 
Therefore, we propose to \textit{support customizable and adaptive representations}, allowing users to display meaningful life-stage appearances while enabling audience-driven selection based on personal memories or conversation topics.

\textit{Knowledge.}
Findings also reveal tensions between identity consistency and incorporation of AI knowledge (in section \ref{ai knowledge}). Proponents believed that extra knowledge could enhance interaction diversity, while opponents worried that it might undermine identity consistency, especially when such knowledge conflicts with their values \cite{hume2000treatise}. In this case, ensuring that any new knowledge aligns with and does not compromise core traits is essential. 
Therefore, we propose to \textit{include careful boundary management when integrating AI knowledge}, suggesting the implementation of pre-defined personal knowledge domains (e.g., education and professional experiences) and value-alignment protocols to prevent identity conflicts.

\textit{Thinking.}
Participants emphasized the importance of agents reflecting both external traits and internal attributes such as thoughts, beliefs, and values \cite{locke1847essay} (in section \ref{simulation level}). This perspective highlights the necessity of designing generative AI-based digital legacy that faithfully represents the personality and inner worlds. Addressing challenges in data collection and organization for capturing inner traits (in section \ref{content}) is a critical step in achieving this goal. 
Therefore, we emphasize the \textit{necessities of innovative data collection methods for inner trait collection}, such as gamified interfaces and context-aware prompts, to capture personal values and thought patterns from their daily life seamlessly and effectively \cite{rabbi2017sara, wilson2005context, dergousoff2015mobile}. 

\textit{Evolution.}
Participants held different views on whether ``AI Afterlife'' should evolve over time (in section \ref{evolution}). While some supported adaptation to new contexts, others raised concerns about preserving core traits to avoid identity confusion \cite{marcia1966development, taylor1989sources}. Evolution can result from both environmental factors and deliberate human intervention in the updating process (in section \ref{update}), requiring careful management to balance adaptability with consistency. 
Therefore, we propose that \textit{evolution should be managed through user-defined update permissions and version control mechanisms}, ensuring adaptability while maintaining core identity markers. 

In summary, designing for continuity in ``AI Afterlife'' involves addressing the complexities of identity across multiple dimensions - appearance, knowledge, thinking, and evolution. These considerations contribute to the broader HCI discourse on digital legacy, providing an ethical and philosophical framework for future afterlife design that prioritizes the preservation of identity while navigating the dynamic nature of generative AI.

\subsection{Design Implications for Balancing Intrusiveness and Support} \label{balance intrusiveness and support}

\ly{
Alleviating grief and creating values for the living are critical factors shaping positive attitudes toward ``AI Afterlife'' as a form of digital legacy (in section \ref{attitudes factors}). While coexistence with this kind of AI-generated agents can provide emotional and practical support for the living, it also raises significant concerns about intrusiveness. Intrusiveness, defined as the degree to which technology disrupts users' attention and emotional well-being \cite{case2015calm}, has been extensively studied in HCI and design research \cite{love2004dealing, kondratova2024cultural, benlian2020mitigating, conti2010malicious, pohl2019charting}. Designing to balance support and intrusiveness involves minimizing disruption, preserving user autonomy, and ensuring emotional well-being, aligning with the goal of creating comforting rather than burdensome digital experiences. Insights from our results highlight three dimensions of intrusiveness - bidirectional interaction, proactivity, and physical space, and emphasize the need for balancing these aspects to ensure meaningful coexistence. 
}

\textit{Bidirectional Interaction Intrusiveness.}
Interactive elements in ``AI Afterlife'' introduce potential for emotional and attentional strain. Bidirectional interactions can be immersive and alleviate loneliness by simulating continued presence, but they also risk prolonging grief through excessive engagement (in section \ref{Interactive and Static}). 
Therefore, we emphasize that designs should \textit{support user control through rule-based mechanisms} that allow individuals to manage interaction intensity and frequency. \textit{Automatic disengagement features should also be incorporated} to prevent undue burden while maintaining meaningful support, such as pausing or ending interactions based on detected emotional distress \cite{slovak2023designing, benke2020chatbot}.

\textit{Proactivity Intrusiveness.}
Proactive behavior of this kind of AI-generated agents, such as sensing and responding autonomously to emotional cues, can lead to discomfort if users feel overwhelmed or that their autonomy is being undermined (in section \ref{reactivity}). 
Therefore, we propose that these \textit{proactive behaviors need customizable settings} that enable users to specify the timing, frequency, context, and type of interactions, ensuring support is delivered only when desired and avoiding unsolicited disruptions to daily life.

\textit{Physical Space Intrusiveness.}
The embodiment of this kind of AI-generated agents, especially in physical forms, poses unique challenges. Participants expressed a preference for virtual agents over physical embodiments due to the emotional and ethical implications (in section \ref{embodiment}). Physical forms can act as persistent reminders of loss, intruding upon emotional spaces and potentially becoming burdensome over time. It is important to respect emotional boundaries and avoid excessive physical presence is essential to ensuring these agents support rather than disrupt.
Therefore, we emphasize that \textit{designs should prioritize virtual forms as the default solution} due to their flexibility and reduced intrusiveness. \textit{Physical embodiments should be limited to specific scenarios} where users express a clear need for tangible interactions. Even in such cases, designs should ensure that \textit{physical forms are easy to manage} which can be adapted or removed to avoid long-term emotional strain.

In summary, the intrusiveness of ``AI Afterlife'' requires careful attention to emotional, ethical, and practical dimensions. By emphasizing autonomy, adaptability, and emotional sensitivity, designers can create AI-generated agent-based digital legacies that coexist harmoniously with the living, providing meaningful support without becoming intrusive.  

%% file: sections/06-Limitations_and_Future_Work.tex
\section{Limitations and Future Work}
While this study provides valuable insights into ``AI Afterlife'' as digital legacy, it is important to acknowledge the limitations stemming from the cultural and social backgrounds of the participants. The research was conducted primarily with individuals from a Chinese cultural context, which inherently shapes their views on life, death, and the afterlife.

In many East Asian cultures, there is a strong emphasis on familial bonds \cite{freedman1961family, qi2016family}, filial piety \cite{hwang1999filial, whyte1997fate}, and the continuity of ancestry \cite{song2015ancestry}. Although participants exhibited some variation, particularly in the context of ongoing social transformation, these cultural values significantly influence how individuals perceive death and the role of AI-generated legacy. For instance, the cultural importance placed on remembrance and honoring ancestors may foster a favorable attitude toward ``AI Afterlife'' as a way to preserve familial connections and legacy. Conversely, the belief in the finality of death and the natural cycle of life may lead to reservations about extensive engagement with AI-generated agents of the deceased. Additionally, in cultures with more individualistic orientations, the emphasis on self-expression may shape attitudes and expectations toward AI-generated agents in unique ways \cite{hunter2008beyond, kim2007express, kim2003choice, inglehart2004individualism}, which contrast with the collectivist tendencies observed in this study, such as prioritizing support for family and alleviating grief.

Given these cultural nuances, the findings of this study may not be fully generalizable to other cultural contexts where beliefs about death and the afterlife differ significantly. However, some key findings, such as the expected interaction designs and concerns, can be applied more broadly. This paper serves as a starting point for understanding human attitudes toward this emerging form of digital legacy. Future research should include a more diverse participant pool from various cultural backgrounds to better explore how different cultural and social values shape perceptions of ``AI Afterlife'' as digital legacy, ensuring it meets the diverse needs and values of users globally.

%% file: sections/07-Conclusion.tex
\section{Conclusion}
This paper presents a qualitative study that investigates perceptions, expectations, and concerns towards ``AI Afterlife'' as digital legacy from the perspective of the individuals being represented by AI-generated agents. 
Through data collection and analysis, We identify factors shaping people's attitudes, their perceived differences compared with the traditional digital legacy, and concerns they might have in real practices. Additionally, We also examine the design aspects throughout the life cycle and interaction process. 
\ly{Furthermore, we situate ``AI Afterlife'' within the context of digital legacy, and delve into design implications for maintaining identity consistency and balancing intrusiveness with supportive interactions in ``AI Afterlife'' as digital legacy.}
In the future, we expect to consider conducting this research in a broader cultural background, so as to explore how different cultural and social values and beliefs influence perceptions and expectations of ``AI Afterlife'' based digital legacy. 